\documentclass{ws06-procs}

\usepackage{graphicx}

\begin{document}

\title{COMBINED SCANNING ELECTRON MICROSCOPE\\
AND MICRO-INFRARED MEASUREMENTS\\
ON INTERPLANETARY DUST PARTICLES}

\author{A. Aronica$^{(1)}$, A. Rotundi$^{(2)}$, G. Ferrini$^{(3)}$,
 E. Palomba$^{(4)}$\\E. Zona$^{(2)}$ and L. Colangeli$^{(1)}$}

\address{$^{(1)}$ INAF - Oss. Astronomico di Capodimonte, Via Moiariello 16,
 80131 Napoli (Italy)\\E-mail: aronica@na.astro.it, colangeli@na.astro.it
\\$^{(2)}$ Dip. Scienze Applicate, Universit$\grave{a}$ Parthenope, Via A. De Gasperi,
 5 - Napoli (Italy)\\E-mail: rotundi@uniparthenope.it, zona@uniparthenope.it
\\$^{(3)}$ Novaetech s.r.l., Piazza Pilastri 18, 80125 Napoli (Italy),
\\E-mail: ferrini@novaetech.it
\\$^{(4)}$ INAF-IFSI, Via del Fosso del Cavaliere 100, 00133 Roma (Italy),
\\E-mail: palomba@ifsi.rm.it}  

\maketitle

\abstracts{
Laboratory characterization of Interplanetary Dust Particles (IDPs) collected in the lower stratosphere represents a concrete analysis of cosmic dust properties which played a fundamental role in the origin and evolution of Solar System. The IDPs were characterized by Field Emission Scanning Electron Microscope (FESEM) analyses and by InfraRed (IR) micro-spectroscopy. We present the FESEM images of six IDPs: three smooth grains, two porous and one a compact sphere. We also show the results of micro-IR transmission measurements on four IDPs that allowed us to identify their spectral class according to the criteria defined by Sandford and Walker~\cite{sa}. Only three of the analyzed particles show IR transmission spectra with a dominant ``silicate absorption feature'' so that they could be assigned to the three IR spectral classes: one has been classified as ``amorphous olivine'', one appears to be a mixture of ``olivines'' and ``pyroxenes'' and one belongs to the ``layer-lattice silicates'' spectral class.}

\section{Introduction}
Cosmic dust which took part in the formation of the protosolar nebula played a fundamental role in the origin and evolution of the Solar System, especially during the planetesimal development. The dust residing in the outer protosolar nebula (30 - 50 AU), formed the minor bodies (i.e. asteroids, comets, {\it Kuiper Belt Objects}) and remained unchanged due to high heliocentric distances and low temperatures~\cite{ri}. As a result of collisions between asteroids and perihelion passage of comets, dust is regenerated as {\it Interplanetary Dust Particles} ({\it IDPs}). Particularly preserved IDPs include interstellar matter and a great amount of information about chemical and mineralogical properties of the primordial dust. For these reasons the study of IDPs origin, chemical composition and physical properties provides an important opportunity to understand the formation and evolution of the Solar System. Dust grains composition is studied in laboratory by InfraRed (IR) micro-spectroscopy of IDPs collected in the lower stratosphere (17-19 km altitude) by high-flying NASA U2 aircrafts. IDPs are analyzed to classify them according to their morphology, physical properties and chemical composition and to identify the link to their parents bodies, mainly comets and asteroids. Many chondritic IDPs are Si-O rich and include carbonaceous materials: the main comets constituents~\cite{br}. Comparisons of IR asteroids astronomical spectra has shown that many IDPs are linked to these bodies although the emission bands near 10 $\mu$m associated with SI-O stretching vibrations is prevented by {\it regolite} which covers the surface of many asteroids~\cite{hi}.

\section{Experimental}
Six stratospheric IDPs were allocated by the NASA-JSC Astromaterials Curation Center (ACC) after selecting from {\it Cosmic Dust Catalog}~\cite{wa}. We designed Special Sample Holders to allow IDPs to be deposited directly at the NASA ACC onto these substrates that are suitable for the combined micro-IR and FESEM analyses without the need for sample manipulation in the laboratory. All bulk particle analyses were non destructive. The morphological analysis was performed with a Stereoscan 360-Leica/Cambridge Field Emission Scanning Electron Microscope (FESEM) at an accelerating voltage of 2.5 kV. The IR spectra were acquired with an IR microscope attached to a FTIR interferometer (mod. Bruker-Equinox 55) in the range 4000 - 600 cm$^{-1}$, at a spectral resolution of 4 cm$^{-1}$.

\section{Results}
\begin{figure}[ht]
\begin{center}
\includegraphics[scale=0.53]{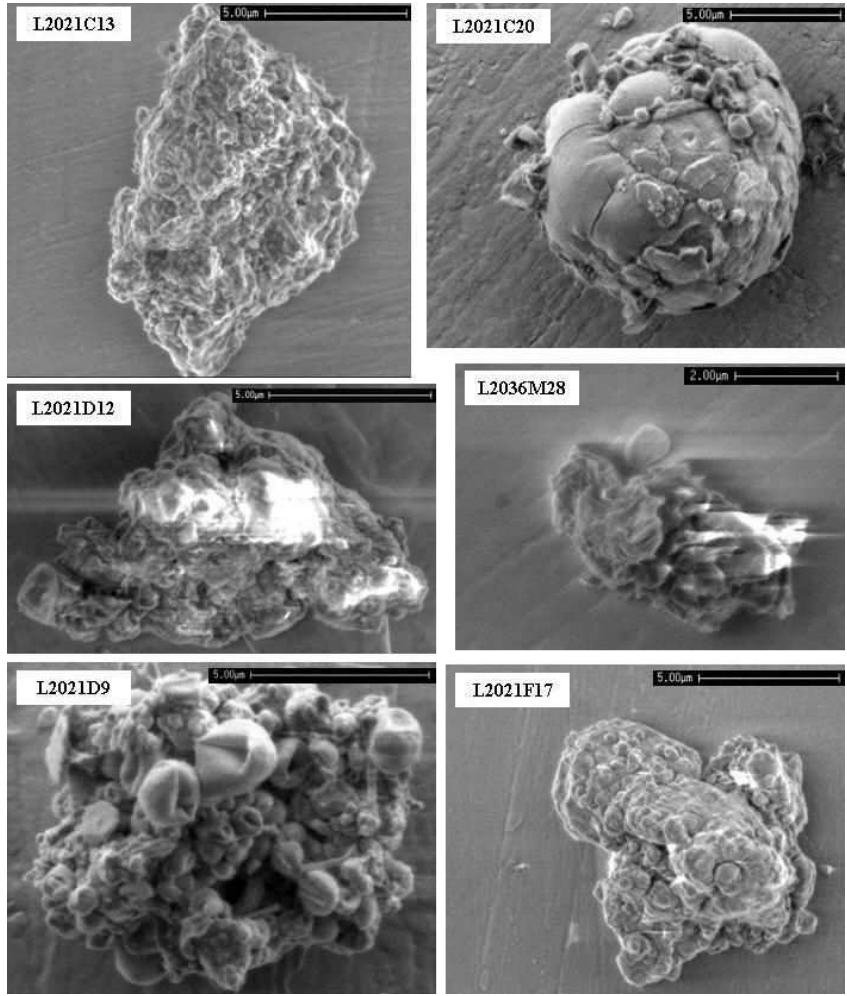}
\caption{FESEM images showing various IDPs morphologies.}\label{FESEM}
\end{center}
\end{figure}

In Figure \ref{FESEM} the IDPs L2021C13, L2021C20, L2021D12, L2036M28 show different morphologies with respect to images contained in the NASA catalog~\cite{wa}. By matching features along the edges of each IDP, we conclude that they were placed upside-down with respect to the NASA catalog images. L2021C13 as imaged in the catalog is an aggregate with an apparent foliation highlighted by platy grain structures on the periphery, while in Figure \ref{FESEM} appear as a low porosity, compact aggregate of stratified submicronic spheroidal grains, with a maximum elongation of 13 $\mu$m. L2021C20 in the catalog is a sphere with a scalloped texture formed by thin platy crystals while our observation (Figure \ref{FESEM}) shows a 13 $\mu$m spherical particle with a smooth core partially covered by micronic and submicronic grains. L2021D12 in the catalog shows up as a very porous particle with an attached about 4 $\mu$m balloon-like feature close to a patch of several fused disk-shaped grains. Instead in Figure \ref{FESEM} the IDP appears as a less porous aggregate of micrometric grains characterized by an irregular form with a max elongation of 13 $\mu$m. The L2036M28 shape and sizes in the catalog image do not match those shown in our FESEM image (Figure \ref{FESEM}) although the stratified texture of this low porosity particle is similar. The particle has probably broken-up during sample manipulation and we observed only a fragment (max elongation of 5 $\mu$m) of the entire particle shown in the catalog. Particles L2021D9 and L2021F17 (both max elongation of 11 $\mu$m) show same morphologies in our FESEM images (Figure \ref{FESEM}) as shown in the catalog. The IDP L2021D9 appears as a heterogeneous aggregate of spherical submicronic grains with high porosity. L2021F17 is a smooth particle covered by a thin granularity and with an irregular form.

\begin{figure}[ht]
\begin{center}
\includegraphics[scale=0.56]{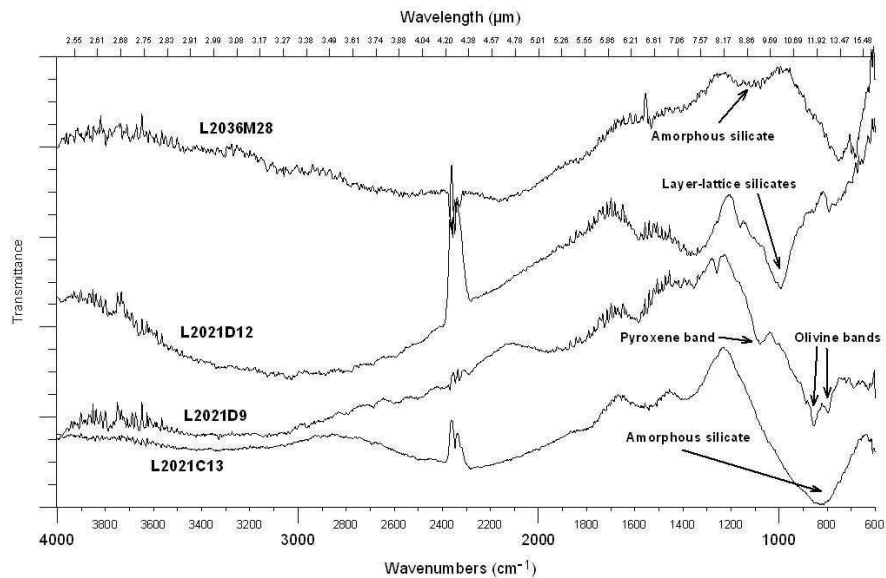}
\caption{Micro-IR spectra of four IDPs.}\label{MicroIR}
\end{center}
\end{figure}

Micro-IR spectra of L2021C13, L2021D9, L2021D12 and L2036M28 are shown in Figure \ref{MicroIR}. We compared them to terrestrial analogs~\cite{sb} and other IDPs IR spectra available in literature~\cite{sa}. The L2021C13 spectrum is smooth and regular and characterized by a wide and strong band at 12 $\mu$m, close to the olivines feature at 11.9 $\mu$m. IR spectra of amorphous materials are more smooth than those of crystalline particles, their strongest bands are shifted towards longer wavelengths and their smallest and narrowest bands are reduced. For these reasons we classify L2021C13 as an ``amorphous olivine''. The L2021D9 spectrum exhibits a sequence of various bands whose the strongest are at 9.3 and 11.7 $\mu$m. These bands are consistent with both Olivine and Pyroxene IR spectral classes. We classified this porous aggregate as a mixture of olivines and pyroxenes in which two anhydrous silicate phases coexist: fayalite and orthopyroxene. The spectrum of L2021D12 exhibits a large and strong band at 10.1 $\mu$m which is characteristic of minerals containing phyllosilicates and therefore we can assign this particle to the ``layer-lattice silicates'' class. The band at 8.6 $\mu$m is due to the presence of iron sulfides (FeS). L2036M28 shows a spectrum characterized by a poor S/N ratio where the 10 $\mu$m ``silicate features'' are not present. For this reason the IR class for this particle can not be established although the large band at 9 $\mu$m could suggest the presence of amorphous silicate.

\section{Conclusions}
We performed IDPs combined Field Emission Scanning Electron Microscope and Micro-InfraRed measurements. These are non-destructive techniques of analyses which allowed us to characterize the particles according to their morphology, physical properties and mineralogical composition. The same combined analyses, on similar substrates, are now applied by our team on the STARDUST samples.

\section*{Acknowledgments}
We thank N. Staiano and S. Inarta for technical support. Work supported by MIUR, ASI, Univ. ``{\it Parthenope}'', Napoli.

\end{document}